\documentclass{aa}
\usepackage{psfig,latexsym,longtable,lscape}

\def\e20{$\times 10^{20}$}
\def\ergsec{erg s$^{-1}$}
\def\ergcmsec{erg cm$^{-2}$ s$^{-1}$}
\def\hnot{$H_\circ$}

\begin{document}

   \thesaurus{3; 
              ( 11.09.1 IC~1262; NGC~6159;
               11.19.
               13.25.2) 
            }

\pagenumbering{arabic}
\title{Peculiar properties of 2 unusually X--ray bright early type galaxies}

\author{G. Trinchieri\inst{1} and W. Pietsch\inst{2}}
\institute{
Osservatorio Astronomico di Brera, via Brera 28, 20121
 Milano Italy
\and 
 Max-Planck-Institut f\"ur extraterrestrische Physik,
              Giessenbachstra\ss e, D-85740 Garching
              Germany
}
   \offprints{G.~Trinchieri}
   \mail{ginevra@brera.mi.astro.it}

   \date{Received date; accepted date}
   \maketitle
 
   \begin{abstract}

Two X--ray bright early type galaxies, selected for their unusually high 
X--ray to optical flux ratio, have been observed at high spatial
resolution with the ROSAT HRI.   Both sources are clearly extended
($> 250$ kpc in radius),  thus
excluding a nuclear origin of the high X--ray emission.  
A small group of galaxies observed around IC~1262 could be the most natural
explanation for the
high X--ray luminosity observed.  NGC~6159 could be similarly explained,
although there are to date no spectroscopic confirmations of a group of
galaxies associated with it.  The X--ray properties of these two sources
are discussed in the context of the properties of groups of galaxies.  
A peculiar very bright feature, in the shape of  an arc, is detected at
the center of the IC~1262 image. 
This could be the signature of a recent merger, or of peculiar high
velocities in the group.

\end{abstract}

\section{Introduction}

The $Einstein$ mission has established that normal galaxies are X--ray
sources with luminosities in the range from $10^{38}$ to $10^{42}$
\ergsec\ (see Fabbiano 1989 for an overview).  
The ROSAT satellite has continued the investigation of
the X--ray properties of these objects, and in particular has provided us
with an X--ray selected, large sample of galaxies obtained from the ROSAT
All Sky  Survey (Voges 1992, Voges et al. 1999,
RASS).  The sample of $\sim$ 600 sources is obtained from the
cross-correlation between the RASS source list and most of the major
optical catalogs of galaxies, and includes objects to within $\sim$
100$''$ from a RASS source.  This sample provides an excellent database
for systematic investigations of the X--ray properties of galaxies, and
in fact some work is well under way (Zimmermann  et al., in
preparation).

To study different aspects of the emission from galaxies, the database
has been sub-divided in subsamples.  One of these deals with objects
peculiar in their higher than expected X--ray luminosity.  It is
composed of very bright X--ray sources, with RASS count rates higher
than 0.1 cts/s.  Care was taken to ensure that all reasonable
explanation for the higher than expected luminosity could be dismissed
($i.e.$ known active nuclei, clusters of galaxies, strongly interacting
systems were not included in the sample).  This sample was the subject
of follow-up ROSAT observations to better investigate the nature of the
X--ray emission.   While in most cases the new data has revealed the
presence of previously unknown nuclear activity (Pietsch et al. 1998), 
several sources stand
out for their unexpected X--ray characteristics.  In particular, the
two early type galaxies discussed here show extended X--ray emission.
While this is not surprising $per~se$, since the hot gas in these
systems is at least coextensive with the stellar light and often
extends further out, 
the X--ray to optical flux ratio (see
Tab.~\ref{parameters}) of these two systems is about one order of
magnitude larger than in most other early-type galaxies studied that
also show very extended X--ray emission.  Emission from a group is a
feasible explanation for IC~1262, the dominant object in a small group
of galaxies (Saglia et al. 1997, and references therein), and could
also apply to NGC~6159, around which several galaxies of similar
optical magnitudes have been identified (''Hypercat: The extragalactic
database\footnote{the on-line version is available at \\
{\tt http://www-obs.univ-lyon1.fr/hypercat/} and mirror 
sites}" and NED\footnote{The NASA/IPAC Extragalactic Database
(NED) is operated by the Jet Propulsion Laboratory, California
Institute of Technology, under contract with the National Aeronautics
and Space Administration. } databases), although no redshift
confirmation is available yet.

 \begin{table}
      \caption{Parameters of IC~1262 and NGC~6159}
         \label{parameters}
         \begin{flushleft}
         \begin{tabular}{lrr}
            \hline
            \noalign{\smallskip}
 IC~1262 & &Ref.  \\
            \noalign{\smallskip}
            \hline
            \noalign{\smallskip}
Type & cD & 1 \\
Assumed distance & 206  Mpc  & 2 \\
& \multicolumn{2}{c}{(hence 1$'\cor60 $~kpc)}  \\
Optical     &  R.A. 17$^{\rm h}$33$^{\rm m}$02\fs1  & 3 \\
position (2000.0) & Dec.
$\phantom{0}$43\degr45\arcmin33\farcs2 & \\
Galactic foreground N$_{\rm H}$ &2.5$\times10^{20}$~cm$^{-2}$ & 4 \\
Optical mag. (B) & 14.05 & 1 \\
X--ray flux & 8.5$\times10^{-12}$ &  5\\
X--ray Lum. & 4.4$\times 10^{43} $&  5  \\
$\rm L_x / L_B$ & 32.5  \\
RASS HR1; HR2 & 0.40; 0.09 & 5\\
            \noalign{\smallskip}
            \hline
            \noalign{\smallskip}
 NGC~6159 & &Ref.  \\
            \noalign{\smallskip}
            \hline
            \noalign{\smallskip}
Type & S0-& 3 \\
Assumed distance & 188 Mpc   & 6  \\
& \multicolumn{2}{c}{(hence 1$'\cor55 $~kpc)}  \\
Optical     &  R.A. 16$^{\rm h}$27$^{\rm m}$25\fs2  & 3 \\
position (2000.0) & Dec.
$\phantom{0}$42\degr40\arcmin47\farcs0 & \\
Galactic foreground N$_{\rm H}$ &1$\times10^{20}$~cm$^{-2}$ & 4 \\
Optical mag. (B) & 15.2  & 3 \\
X--ray flux & 1.7$\times10^{-12}$ &  5\\
X--ray Lum. & 7.5$\times 10^{42} $&  5  \\
$\rm L_x / L_B$ & 32.2  \\
RASS HR1; HR2 & 0.90; 0.40& 5\\
            \noalign{\smallskip}
            \hline
            \noalign{\smallskip}

         \end{tabular}
         \end{flushleft}
{
References: (1) Saglia et al. 1997 ;
(2)  from the group distance, Wegner et al. 1996, 1999, assuming  
\hnot = 50 km s$^{-1}$ Mpc$^{-1}$ ;
(3) from NED ;
(4) Dickey \& Lockman 1990 ;
(5) Unabsorbed X--ray fluxes and luminosities in the 0.1-2 keV band in
c.g.s. units; 
Hardness ratios HR1 and HR2 in two sets of bands, as defined in RASS;
see text 
(6) LEDA database and  Bischoff et al. 1999, assuming
\hnot = 50 km s$^{-1}$ Mpc$^{-1}$
}
   \end{table}

In the present paper we present a detailed study of the morphology of
the X--ray emission of these sources and speculate on their nature.

\section{Results of the ROSAT data analysis}

Both sources were first observed in follow-up snap shot ROSAT HRI
observations of $\sim 5$ and $\sim 7$ ks for IC~1262 and NGC~6159
respectively.  These were used to confirm the identifications and
establish the extended nature of the
sources but could not be used for a detailed study of the morphology of
the emission.  Therefore 2 longer observations, of 26 and 56 ks
respectively were obtained in AO7 with the ROSAT HRI.   A full
description of the ROSAT satellite and of the instrument can be found
in Tr\"umper (1983), Tr\"umper et al. (1991) and Zombeck et al. (1995).
The data analysis was done mostly with the {\it xray} package in
\verb+IRAF+.

\begin{figure*}
\unitlength1.0cm
\begin{picture}(18,9.0)
\thicklines
\put(0,0.0){
\begin{picture}(18,9.0)
\resizebox{18cm}{!}{
\psfig{figure=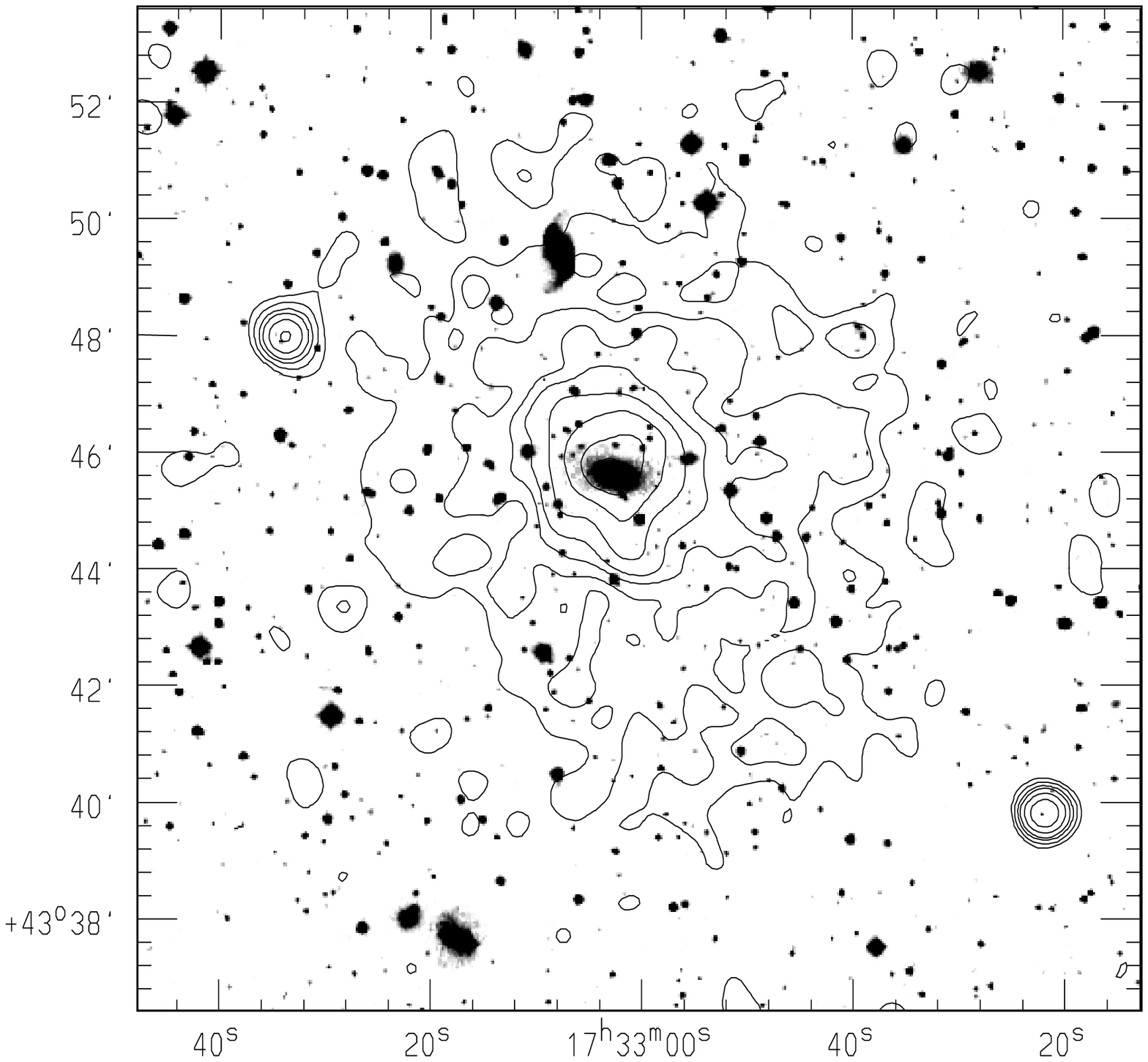,width=18cm,clip=}
\psfig{figure=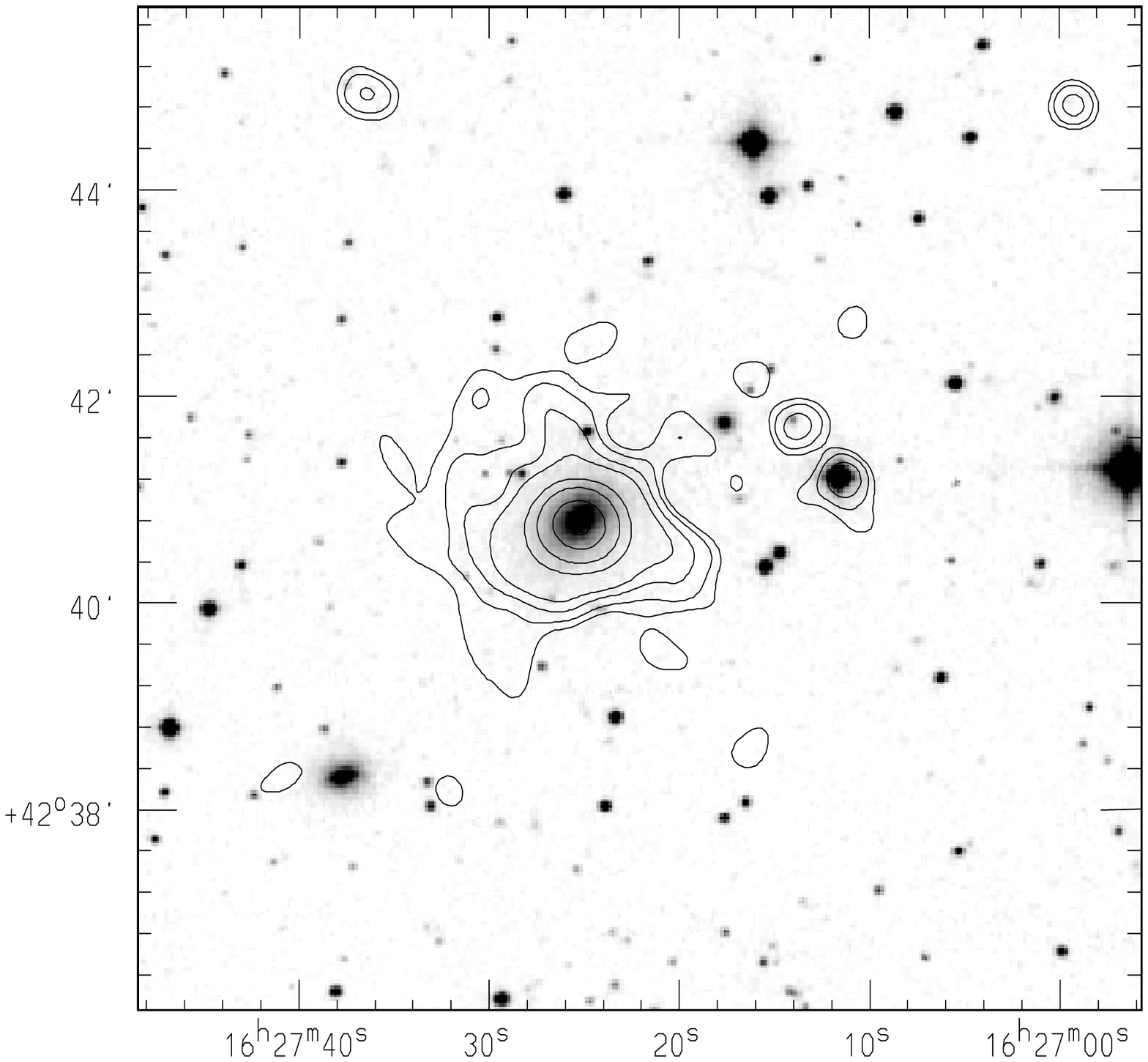,width=18cm,clip=}
}
\end{picture}}
\put(4.0,8.2){
IC~1262}
\put(13,8.2){ 
NGC~6159 
}
\end{picture}
\caption{Isointensity contour plot of the X--ray emission in IC~1262
(LEFT) and NGC~6159 (RIGHT) superposed on the DSS images.  The data have
been smoothed with a Gaussian function with $\sigma$=15$''$ (IC~1262), and
$\sigma$=10$''$ (NGC~6159).
Contour levels are: 0.04, 0.05, 0.06, 0.08, 0.1, 0.14, 0.2 cnt
arcsec$^{-2}$ (IC~1262) and  
0.1, 0.11, 0.13, 0.16, 0.24, 0.32, 0.48 cnt arcsec$^{-2}$
(NGC~6159)}

\label{map}
\end{figure*}

Figure~\ref{map} presents the isocontours of a smoothed X--ray image
overlayed onto the  Digitized Sky Survey plate\footnote{The Digitized
Sky Survey was produced at the Space Telescope Science Institute
(STScI) under U.S. Government grant NAG W-2166.} for the two galaxies
studied here.  

To produce the X--ray map, we have first made a selection in the Pulse Height
Analyzer (PHA) channels.  While the HRI does not have a full spectral
calibration, the data retain the information on the incoming PHA
channel, therefore a comparison between the spectral photon distribution of 
the background and the source (in PHA space) is useful to eliminate
background-dominated channels and improve on the signal-to-noise.  
However, care should be taken when the source is extended, since the
gain is not constant across the detector.  This has the effect of
shifting the distribution of the photons, so that an estimate of the
background in a region of significantly different gain could be not
representative of the source region.  Therefore the selection in PHA
channels should be such as to maximize the signal-to-noise ratio while
retaining a large enough range in PHA space so that gain changes across
the detector do not significantly affect the data analysis. In most
cases, excluding PHA channels $>$ 10 is enough to considerably reduce
the background contribution without introducing significant biases in
the local background estimation (see also the discussion on the HRI
background in ``The ROSAT High Resolution Imager (HRI) Calibration 
Report"\footnote{available on line at {\tt
http://hea-www.harvard.edu/rosat/ \\ rsdc\_www/hricalrep.html}}).  The
data in PHA space 1-10 have then been smoothed with Gaussian functions
for graphical display and for a visual understanding of possible
features in the data.

We have then determined the extension of the emission as a whole, and selected
a region free of emission as the background.   Since the background in  the HRI
is relatively flat across the detector, determining it in an annulus around
the source and applying the resulting count rate to regions at smaller off-axis
angles is a reasonable approximation.  The radial distribution of the total
emission indicates excess emission within radii  r$\sim 7'-8'$ and r$\sim
4'-5'$ for IC~1262 and NGC~6159 respectively.  We have therefore assumed a
total extent of r=7$'$ and r=5$'$ respectively and estimated the
background level in the adjacent annuli (7$'-11'$ and $5'-7'$).   A more
detailed analysis customized for each source was then followed, as described in
the next two sections.

\subsection{IC~1262}

\begin{figure}
\unitlength1.0cm
\begin{picture}(8.5,8.5)
\thicklines
\put(0,0.0){
\begin{picture}(8.5,8.5)
\psfig{figure=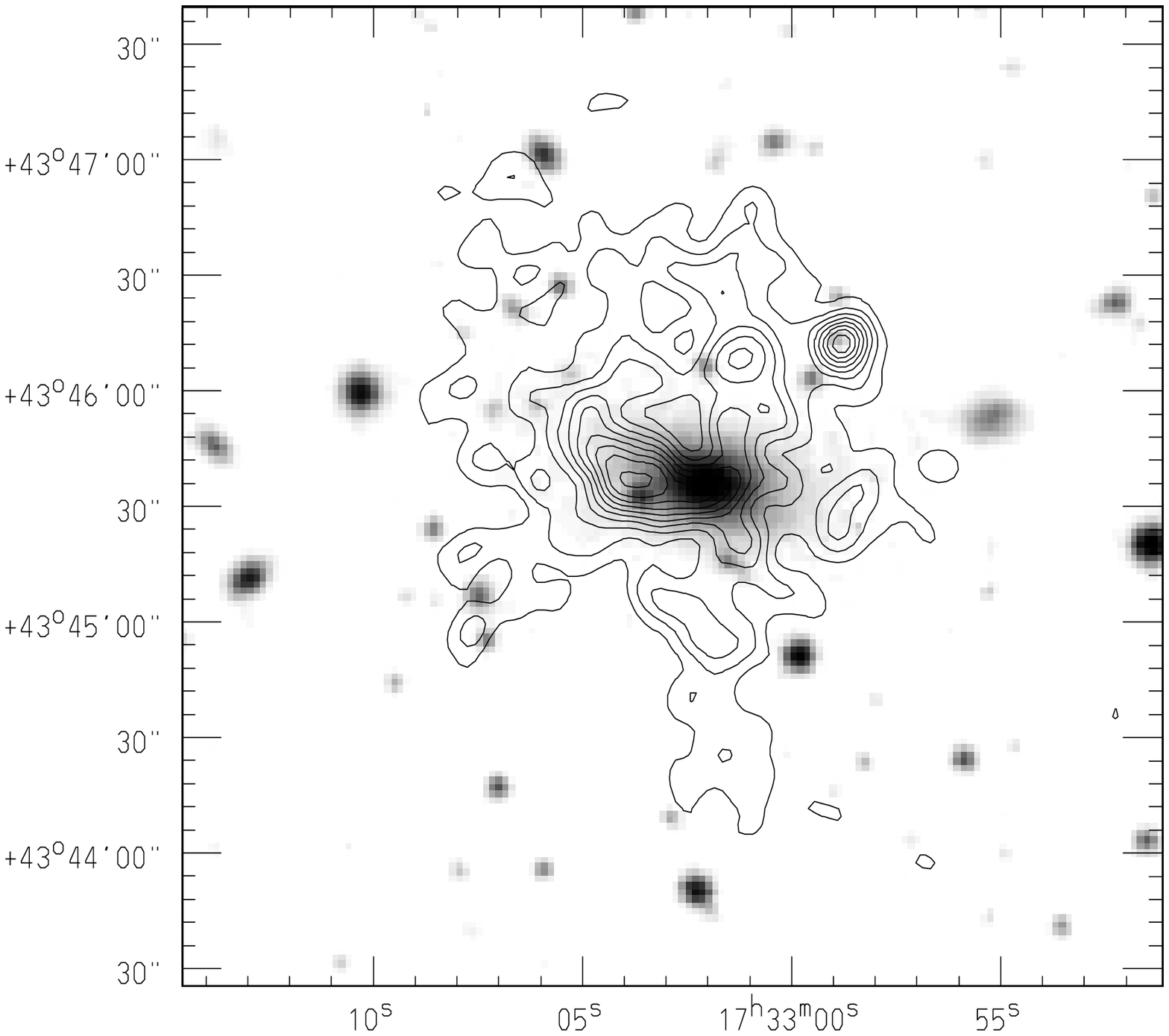,width=9cm,clip=}
\end{picture}}
\put(4.0,8.2){
IC~1262}
\end{picture}
\caption{Isointensity contours of the central region of the HRI image of 
IC~1262 on the Palomar DSS image.  The X--ray data have been smoothed with a
Gaussian function of $\sigma = 4''$. Contour levels start at
0.125 cnt arcsec$^{-2}$ and increase by $\sim 0.038$.}
\label{map-ic}
\end{figure}

\begin{figure}
\psfig{figure=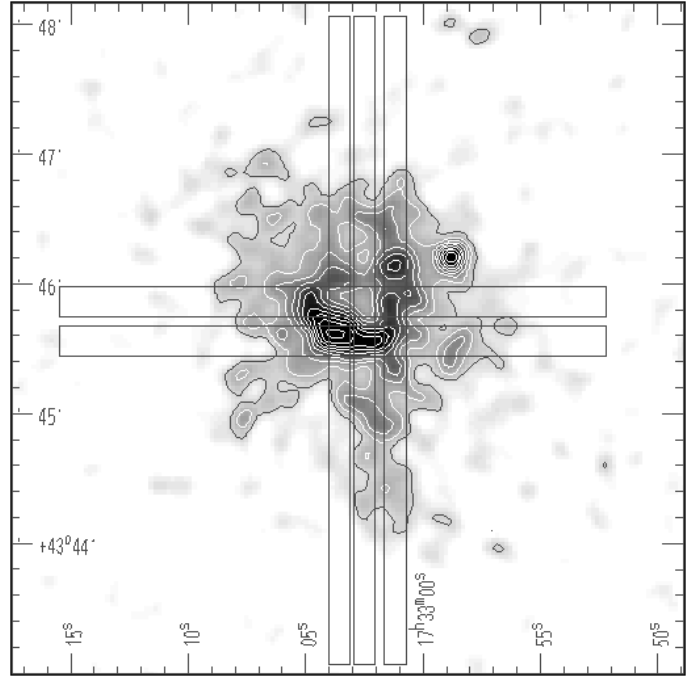,width=9cm,clip=}
\caption{X--ray image of the central part of IC~1262 (Same smoothing and
contour levels as  Fig.~\ref{map-ic}).  
The boxes drawn indicate the areas from which projections along the
longer dimension are obtained and plotted in Fig.~\ref{icew} and
Fig.~\ref{icns}.}
\label{icdet}
\end{figure}

\begin{figure}
\psfig{figure=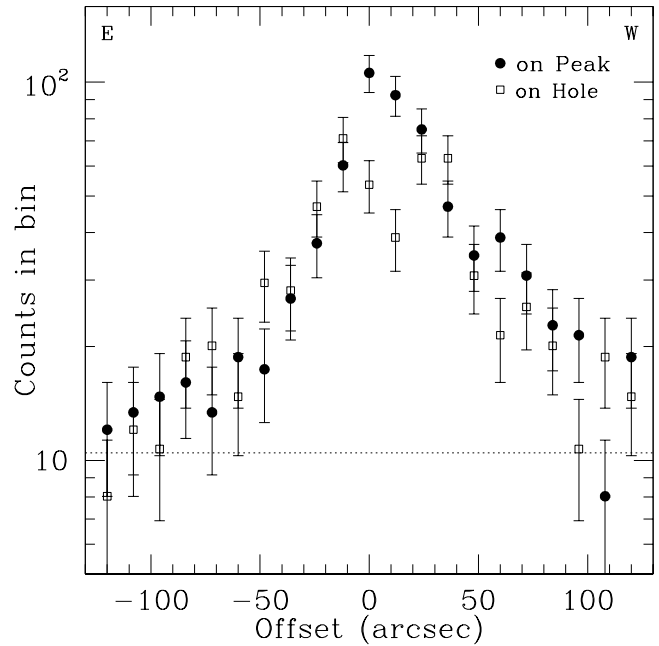,width=9cm,clip=}
\caption{Count distribution along the major axis of two narrow boxes
($14'' \times 252''$) oriented
E-W (see Fig.~\ref{icdet}), and  centered on the X--ray peak and on the
depression N of it. Bins are $\sim 12''$ width. 
The dotted line is an estimate of the average $\Sigma _x$ E and W of the
``on peak" projection. 
}
\label{icew}
\end{figure}

\begin{table*}
\caption{Net counts from different features observed at the center of
the IC~1262 X--ray image}
\label{features}
\begin{tabular}{lrrrrrll}
\hline
Name&Net Counts&Error&\multicolumn{1}{c}{Area}&Sur. Brigh.&Error&Region \\
&&&\multicolumn{1}{c}{(arcsec$^2$)}&\multicolumn{2}{c}{(cnt/arcsec$^2$)}
&Description \\
\hline
&&&&\\
Arc&644.74&28&2440&0.264&0.011 &Polygon surrounding \\
&&&&&& the high surface \\
&&&&&& brightness structure \\
N of arc& 40.64&8.5&384 &0.106 &0.022 & Polygon surrounding the \\
&&&&&& low surface brightness region \\
&&&&&& between the ``arms" of the arc \\
PS&69.33 &10 &312&0.222&0.03& 10$''$ radius circle \\  
Southern tail & 234.4 &18 & 2076 &0.1228 & 0.0087& Narrow 24$''
\times 84''$ \\
&&&&&& box aligned with the \\
&&&&&& extension to the S\\
Inner 1$'$& 893.37& 35.6&8172&0.109 & 0.004 & 1$'$ radius circle with the\\
&&&&&& exclusion of all previous \\
&&&&&& regions centered at \\
&&&&&& 17$^h$33$^m$02.6$^s$ 43\degr45$'$55$\farcs$2 \\
1$'-2'$ Annulus& 1013.81& 43 & 20112 & 0.050& 0.002 & Adjacent
concentric  annulus \\
&&&&&& with the exclusion of \\
&&&&&& all previous regions \\
\hline
\end{tabular} 

Notes:\\
Net counts above the average field background estimated in the 7$'-11'$
annulus around the source (see text). \\
\end{table*}

We have looked in more detail at the morphology of the
X--ray image of IC~1262 and at how it compares to the optical image.   In
Fig.~\ref{map-ic},  the smaller smoothing function applied to the data
shows more clearly the  complexity of the source.   A  point-like
source is visible to the NW, centered onto a faint optical
counterpart, most likely an interloper.  A large fraction of the rest
of the photons is located in an arc-like structure, peaked to the east
of the galaxy, just to the north of another optical source.

We have tried to evaluate the significance of these structures.  
We have defined regions (like the enhanced emission in the arc-like
structure, the point-like source to the NW, the tail to the S) and
compared the net counts there to those of the region immediately N of
it, and to the surrounding area.  Tab.~\ref{features} summarizes the
results.

\begin{figure*}
\psfig{figure=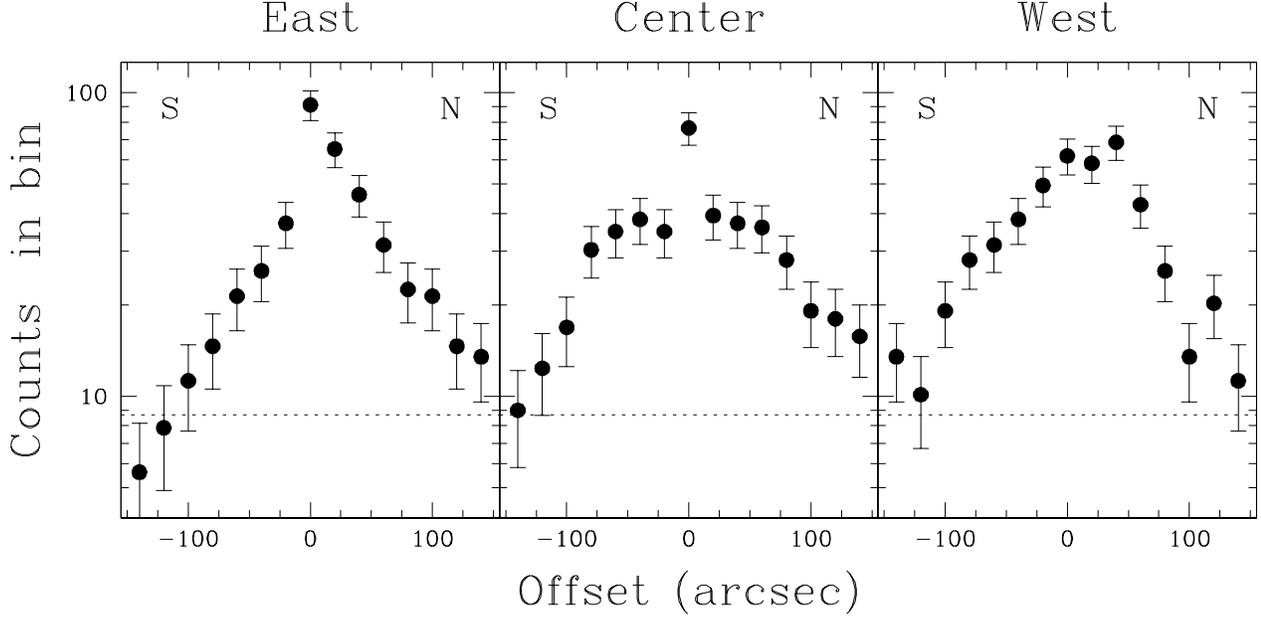,width=18cm,clip=}
\caption{Count distribution along the major axis of  three narrow boxes
($10'' \times 300''$) oriented N-S (see Fig.~\ref{icdet}). 
Counts are binned in  $\sim 20''$ width bins.
The dotted line is an estimate of the average $\Sigma _x$ N and S of the
``center" projection. 
}
\label{icns}
\end{figure*}

As can be seen from the table, the arc-feature has a significantly
higher surface brightness than the surrounding emission. 
The region just north of the
arc, and surrounded by it,  is 
consistent with the average residual emission in the inner 1$'$ radius
circle, and higher than the external adjacent annulus.
If we assume the average surface brightness in the inner 1$'$ circle outside
the arc as
a reasonable estimate of the local background, there are $\sim 390\pm 27$ 
excess counts in the arc.  These can further be divided into a higher  
and a lower surface brightness region, with a factor of $\sim 2$ in
surface brightness contrast (see also Fig.~\ref{icew} and
Fig.~\ref{icns}).
The peak of the arc is to the East of the galaxy, just north of 
a separate optical object visible on the plate.  We will discuss the
possible relation between this optical source and the arc feature 
later.  The southern extension, also clearly visible in
Fig.~\ref{map}, represents a significant excess over the average
surface brightness in the 1$'-2'$ annulus.  It can also be further
separated in a higher and lower surface brightness regions, with again a
factor of $\sim 2$ contrast, that makes it a significant enhancement
over the average emission in the 1$'$ radius circle. 

The reality of the structures are also illustrated by the plots of the
photon distributions along different projections in narrow boxes
aligned with the different structures (Fig.~\ref{icew} and 
Fig.~\ref{icns}).   The positions of these boxes
are superposed onto the X--ray contours  in
Fig.~\ref{icdet} for easy reference.  

In Fig.~\ref{icew}, projections along the two boxes oriented E-W,
and aligned with the high surface brightness in the arc (labeled ``on 
Peak" in the figure) and with the ``hole" north of the arc, 
are compared.  In the central $\pm 30''$ from the box center, a
significant contrast in surface brightness is
evident. The projection along the arc (E to W, filled symbols) 
shows a fast rise onto a single peak, followed by a somewhat shallower
decline.  The  projection in the box North of it 
(open symbols) clearly marks the
crossing of the two arms of the arc (at $\sim -10''$ and $\sim +25''$
from the box center), and shows significantly lower emission in
between.    In both profiles a relatively smooth surface brightness
profiles is apparent outside of $\pm 50''$ offset.

Figure~\ref{icns} also indicates a rather complex photon distribution.  
The Easternmost box shows a pronounced peak and a relatively sharp 
fall off of the surface brightness steeper to the south than to the north of
it, in accordance with the presence of an extension of the arc to the N.  
The projection in the central panel indicates a relatively flat
plateau $\sim 160''$ long, with a very high peak in the middle
corresponding to the intersection with the arc.  
No sharp peak is visible in the projection along the Westernmost box,
and a shallow increase in surface brightness visible from roughly
$-100''$ (the ``tail") is followed by a flatter distribution to $+50''$
(the western arm of the arc with the secondary peak), 
and a faster decline outside. 

\begin{figure}
\psfig{figure=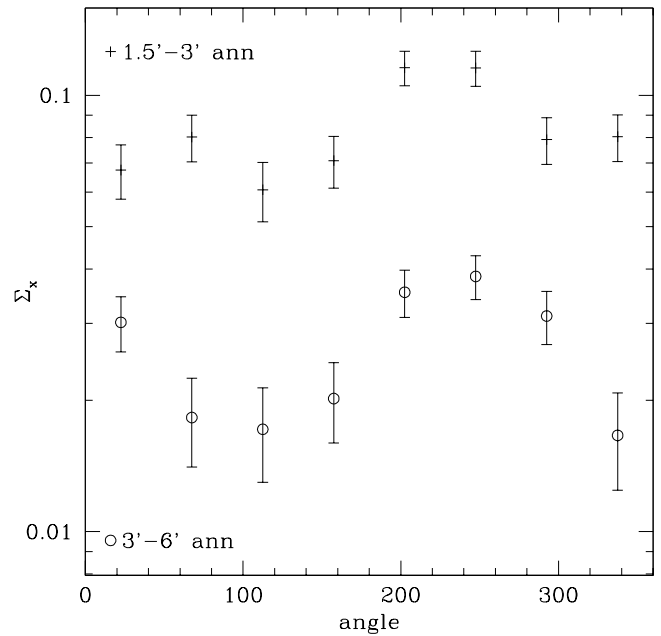,width=9cm,clip=}
\caption{Azimuthal distribution of the total emission from IC~1262 in
two adjacent concentric annuli.  
}
\label{iazim}
\end{figure}

At larger radii, the emission appears smoother and relatively symmetric,
although a significant modulation in the azimuthal photon distribution
can be clearly seen in the plot of Fig.~\ref{iazim}.  This points to an
extension to the S-SW, and a possible compression of the emission in the
NNE sector. 
Therefore, an azimuthally averaged radial profile, that fails to
represent the data at small radii, 
is a rather poor representation of the true radial distribution
even at larger radii.

\begin{figure}
\psfig{figure=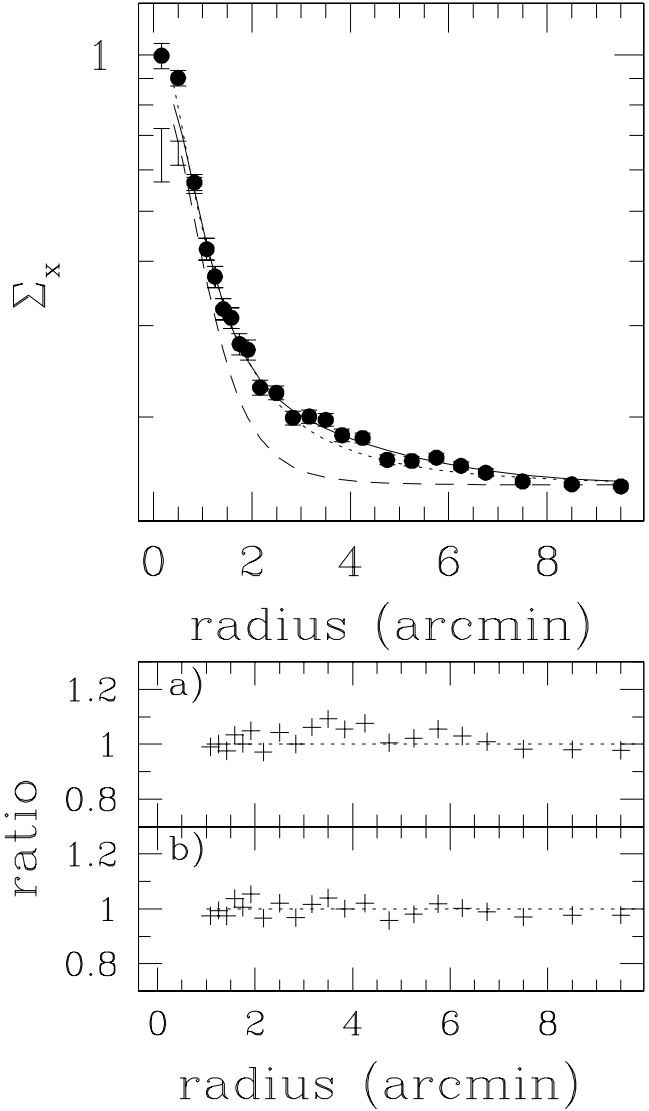,width=9cm,clip=}
\caption{Surface brightness profile of the total emission centered on
IC~1262.  {\bf Top panel}: Azimuthally averaged data with statistical
errors (filled dots). The models plus the field background discussed
in the text are plotted: 
{\it model a} r$_c(1) = 80''$ and r$_c(2) = 400''$ and
$\beta=1$ (solid line);
and {\it model b}: r$_c(1) = 38''$, $\beta=0.5$ (dotted  line).
The short-dashed line represents a King profile with r$_c = 80''$,
$\beta=1$ (one component of {\it model a}).  {\bf 
Bottom panels}: ratio of the
azimuthally averaged data to the models. Top: data vs model
{\bf a};  Bottom: data vs model {\bf b}.
}
\label{iprof}
\end{figure}

We have nonetheless produced the azimuthally averaged profile of
Fig.~\ref{iprof} for comparison with other groups. The structure at the
center (the arc, see Tab.~\ref{features}) has been either included or
taken out from the profile.  We therefore expect that the surface
brightness of the ``unperturbed" component in the innermost bins (of
20$''$ radius each) is between the two profiles.  We have tried to
parameterize the emission with a ``King-type" profile $
 \Sigma _x \propto (1+ ({r\over r_c} )^2)^{-3\beta+0.5} $ generally a
good representation of the data,  excluding the innermost $\le
1'$,  where the arc structure confuses matters.  Mulchaey \& Zabludoff
(1998)  have parameterized  the data for several nearby groups with a
combination of 2 ``King-type" models.  If we assume their best fit
$\beta \sim 0.8-1$, we find that a model with r$_c(1) \sim 80''$ and
r$_c(2) = 400''$ and $\beta=1$ reproduces the azimuthally averaged data
reasonably well (see Fig.~\ref{iprof} ).  A $\beta$=0.8 would require
core radii r$_c(1) \sim 1\farcm5$ and r$_c(2)\sim 8'$.  However, a
single ``King" model with $\beta$=0.5 and r$_c(1) \sim 38''$ gives an
equivalently good representation of the azimuthal profile outside
$\ga 1'$ (see Fig.~\ref{iprof} ).

\subsection{NGC~6159} 

Figure~\ref{map} (right panel) shows the X--ray isointensity contours
of the inner part of the NGC~6159 field overlayed onto the Palomar DSS. 
The figure shows a relatively regular, extended emission centered onto
the optical galaxy, and four separate sources to the N and to the W of
it.  All of these sources  are coincident with optical counterparts, one
with a bright star and the other three with fainter point-like objects
(see appendix).
This ensures that the absolute coordinate positions are correct. 

\begin{figure*}
\unitlength1.0cm
\begin{picture}(18,15.0)
\thicklines
\put(0,0.0){
\begin{picture}(18.00,15)
\psfig{figure=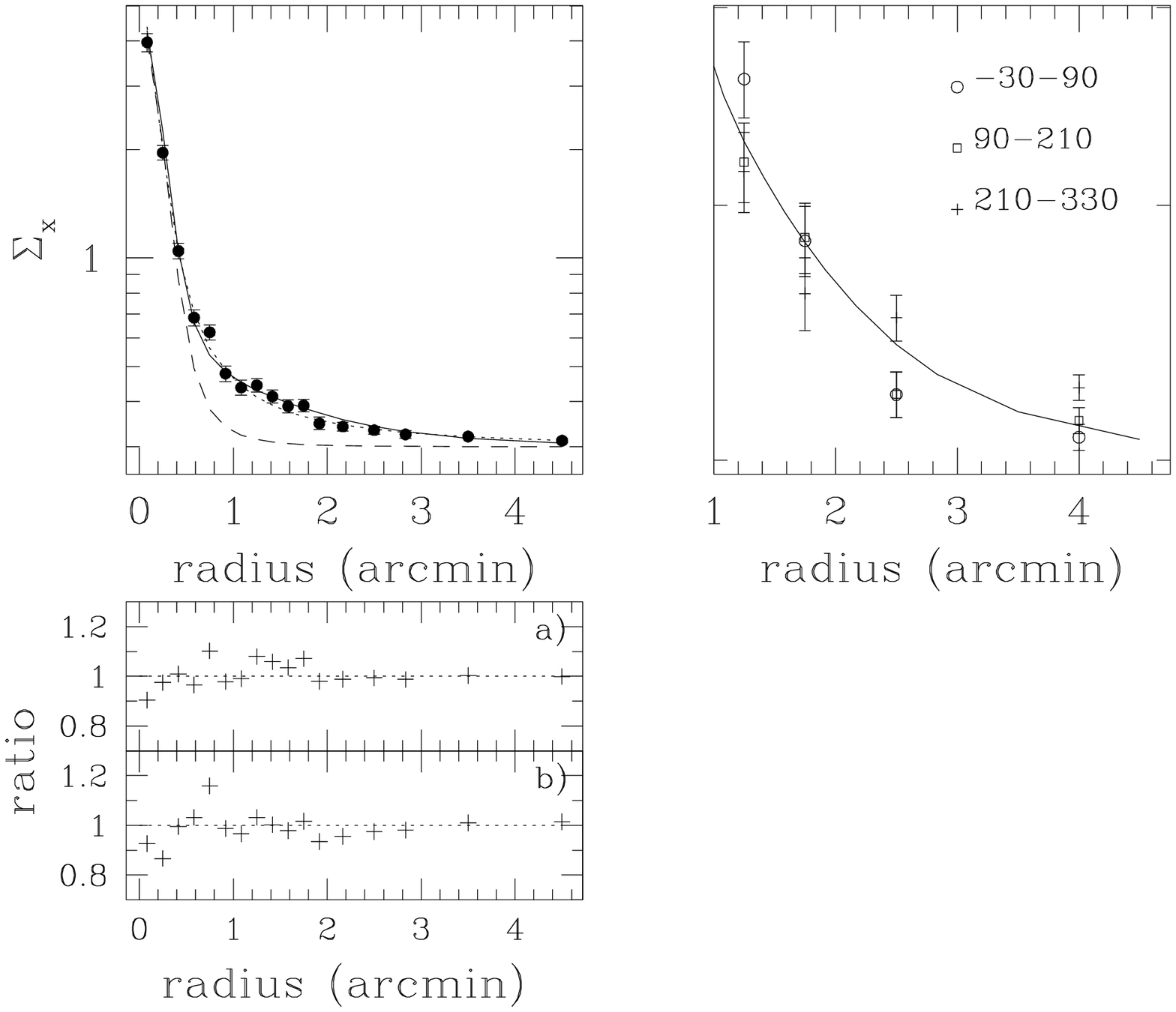,width=18cm,clip=}
\end{picture}}
\put(9.5,2.6){
\parbox[]{8.3cm}{
\caption{
Surface brightness profile of the total emission centered on
NGC~6159. {\bf Top left}: Azimuthally averaged data with statistical
errors (filled dots). The sum of 2 King models plus the field background are 
plotted: 
{\bf model a} r$_c(1) = 20''$, r$_c(2) = 160''$ and $\beta=1$ (solid line); 
and {\bf model b}: r$_c(1) = 8''$, $\beta=0.5$ and r$_c(2) = 300''$,
$\beta=1$ (dotted line). 
The dashed line represents a single King profile with r$_c = 20''$,
$\beta=1$ (one component of {\it model a}).  {\bf Top
right}: Different
angular sectors are compared to the combination of the 2 King profiles
(solid line, same as left panel).     Angles are measured
counterclockwise from North.  {\bf Bottom}: ratio of the
azimuthally averaged data to the models. Top: data vs model
{\bf a};  Bottom: data vs model {\bf b}.  
}
\label{nprof}
}
}
\end{picture}
\end{figure*}

With different smoothing functions applied to the data, we have searched
for substructures in the morphology, of the
same kind as those in IC~1262.   Within the present data, no
significant substructures at small radii are detected.  As shown by 
Fig.~\ref{map}, however, the source does not appear azimuthally
symmetric.   To assess the significance of the visual impression, we
have derived the radial profile in 3 angular sectors (see
Fig.~\ref{nprof}).  Outside $\sim 1'$, the emission is distributed
differently at different angles:  it appears to be
stronger and less extended in the NW region ($\le 3'$), and more
extended to the SW ($\ga 4'$).  
As for IC~1262, the azimuthally averaged profile therefore only
approximates the photon radial distribution.  Nonetheless, we have
compared it with ``King-type" model(s).  
For this galaxy we cannot find a set of core radius and $\beta$ to
represent the whole profile, and at least 2 models are required, albeit
with very poor constraints on the parameter space:  core radii of r$_c \sim
20''$ and r$_c \sim 160''$ and $\beta=1$ plus the field background are
used to draw the solid line in Fig.~\ref{nprof}.  However, a
combination of r$_c(1)=8''$, $\beta(1)=0.5$, and r$_c(2)=300''$ and
$\beta(2)=1$ also represent the data equally well (dotted line in
Fig.~\ref{nprof}).

\subsection{Fluxes and Luminosities}

Table~\ref{parameters} summarizes the optical and X--ray data for the
two galaxies discussed here.  No reliable spectral information can be
obtained from the HRI data to convert from observed count rates to
X--ray fluxes.  However, we can use the spectral
information contained in the ROSAT PSPC survey data.  Two
hardness ratios are derived for each RASS source, defined as the
ratio of the counts in different energy bands within the ROSAT
band: HR1 = (h--s)/(h+s) and HR2 = (h2--h1)/(h2+h1); where ``h'' is the
0.5--2.0~keV band, ``s" =0.1--0.4 keV, ``h1"= 0.5--0.9~keV and ``h2" =
0.9--2.0~keV.  While these ratios cannot distinguish between models,
they can be used
to estimate crude spectral parameters within a given spectral model.
Fig. 1 in Pietsch, Trinchieri and Vogler (1998) shows the distribution
of hardness ratios for different models, as temperatures (or power law
indexes) and low energy absorptions vary within reasonable values.  In
the framework of a thin plasma model, that is used for the average
spectrum of high X--ray luminosity early type galaxies or small groups, the
HR1 and HR2 of either galaxies are consistent with low 
absorption and a temperature of 1-2 keV.  We have therefore adopted a 1
keV temperature, and the line-of-sight column density
for the counts to flux conversion.  We also note that in the relatively
narrow ROSAT band the conversion factors are not very sensitive to
the models assumed, provided that the low energy absorption is known
and/or small.  Large amounts of unaccounted for absorption could cause
drastic underestimates of the emitted flux.

Total fluxes and luminosities ($i.e.$ within a radius of $7'$ and 5$'$
for IC~1262 and NGC~6159 respectively) are given in
Table~\ref{parameters}.  Luminosities for the single components of
IC~1262 can be rescaled from the net counts listed in
Tab.~\ref{features}, using a conversion factor of 100 counts = 1.3
$\times 10^{-13}$ \ergcmsec, appropriate for the 1 keV plasma spectrum
and line-of-sight Galactic absorption.

The arc in IC~1262 is also a very bright source of emission.  If
we consider it as a local enhancement, above the average surface
brightness in the 1$'$ radius inner circle, it has a luminosity L$_X
\sim 2.6 \times 10^{42}$ erg s$^{-1}$, for the same assumption of spectral
parameters.   If we include the tail, the luminosity of the whole
structure is L$_X \sim 3.2  \times 10^{42}$ erg s$^{-1}$.

\section{Discussion}

The X--ray luminosities of both galaxies are significantly higher than
10$^{42}$ erg s$^{-1}$, indicating intrinsically bright sources, and  
their $\rm L_x /L_B$ ratios are also high (see Table~\ref{parameters}), in
particular if compared to values  for normal galaxies (Fabbiano, Kim  and
Trinchieri 1992; Beuing et al. 1999).  Their X--ray
luminosities are  instead comparable to those 
of optically or X-ray selected groups of galaxies (Mulchaey et
al. 1996; Ponman et al. 1996; Burns et al. 1996; Mahadavi et al. 1997 
and references therein). 

The presence of a group of mostly early type galaxies around IC~1262
(Wegner et al. 1996) indeed indicates that the X--ray emission should be
associated with the potential of the group rather than with that of the
galaxy only.  The total extent, $\ga  400 $ kpc in radius, also suggests a
much larger potential than usually associated with galaxies.  The
complex morphology at the center, however, is probably related to 
the central galaxy (see \S~\ref{thearc}).

The optical spectrum of NGC~6159 indicates a Sey2/Liner nuclear source
(Bischoff et al. 1999).  However, the X-ray  
source is clearly extended, with no indication of a central point source
embedded in a more extended component.  Therefore, we do not expect an
association between the X--ray emission and the galaxy's nucleus. 
There is at the present time no confirmation that a group exists around
NGC~6159.  However, Ramella et al. (1997) have identified 
several groups in the area (within a few degrees from NGC~6159) 
in a very large structure at the same average redshift (see data in the 
Huchra CfA Redshift Catalog 
available on line at {\tt http://heasarc.gsfc.nasa.gov}).  
Although NGC~6159 itself does not have an
entry in the catalog, and the only velocities available for galaxies
within 1$^\circ$ radius are significantly higher, there are
7 galaxies $< 1.5$ degree south of it with measured velocities in the
range 8200-9600 km s$^{-1}$  and 27 within 2 degrees with v$\le 10000$  km
s$^{-1}$
(for comparison, NGC~6159 has a velocity
of $v$= 9480$\pm$150, Bischoff et al. 1999;  or $v$= 9380.0, Huchra et
al. 1983).  Since there are several fainter galaxies close to NGC~6159
on the sky, for which no redshift information is available,  we cannot
exclude that NGC~6159 resides in a rich, though loose, environment.

The extent of the emission ($\ge 250$ kpc in radius)
suggests the presence of a local condensation of matter, since it is a
signature of a larger potential than expected from a single galaxy.
Vikhlinin et al. (1999) have recently suggested that a new
class of objects exists, namely the Over-Luminous Early type Galaxies
(OLEG), large dark mass concentrations associated with a single
isolated early-type galaxy, that can be discovered through the high
X--ray luminosity and X--ray-to-optical flux ratio.   Equivalently,
``failed groups" or merger remnants could be discovered from their
extended X--ray halos associated with a galaxy with optical properties
that would not qualify it as a ``group" (NGC~1132 is an example,  
Mulchaey \& Zabludoff 1999).   While NGC~6159
shares some of the properties of these systems (high X--ray luminosity
relative to the optical luminosity, large extent, apparent isolation),
these are not as extreme as those of
OLEGs, and NGC~6159's apparent isolation is not confirmed
by a thorough optical study as in the case of NGC~1132.  
We therefore will assume that the high X--ray luminosity and extent of
this system are due to the presence of a poor group of galaxies.

We can distinguish three main components contributing to the emission
in IC~1262:  an  extended component, a point-like source not at the
galaxy's nucleus, and a complex central structure.  
Only the first component is detected in NGC~6159.

\subsection{Emission from a group of galaxies}

X--ray observations of several compact and poor groups have
already shown the presence of extended emission in these systems, associated 
with a hot intergalactic medium with typical temperatures of $\sim$ 1-2
keV, luminosities up to $10^{43}$ erg s$^{-1}$ and total gas content
comparable to or lower than that in the luminous mass (see Mulchaey et
al. 1996; Ponman et al. 1996, Burns et al. 1996; Mahadavi et al. 1997 
and references therein).    

As already discussed, both the X--ray luminosities and the 
dimensions  of these systems are
appropriate for groups.  The gas density and mass are also consistent
with those of similar systems.  Assuming a simple spherical
geometry to the hot gas distribution, and a temperature of 1 keV (hence
a cooling coefficient  $\Lambda(10^7)$ = $5 \times 10^{-23}$, Raymond
et al.  1976, see Nulsen et al. 1984),  the average gas density for
IC~1262 is n$_e$ $\sim 3.2 \times 10^{-4}$, and the total gas mass
M$_{gas} \sim $ 3$\times 10^{12}$ M$_\odot$, calculated from the luminosity 
of the whole source (within a radius of $\sim 400$ kpc)
with the exclusion of the contributions
from the interlopers and the arc feature.  For NGC~6159,  the average
n$_e$ $\sim 2.7\times 10^{-4}$, and the total gas mass M$_{gas} \sim $
7$\times 10^{11}$ M$_\odot$ within a $\sim$ 250
kpc radius.   

A recent careful analysis of the morphology of several nearby groups
(Mulchaey \& Zabludoff 1998) has shown that, in most cases,  
the extended X--ray sources associated with them can
be modeled with two King-type functions, with
$\beta \sim 0.8-1$ and different size core radii.  Mulchaey \&
Zabludoff propose that the  smaller core radius model is associated
with the galaxy that is found at the center of the emission.

As shown in the previous sections, the data for these two sources also
indicate that a single King model does not parameterize well the full
range of the radial photon distribution, but that a combination of 
models could be used for this purpose.  In spite of the higher
resolution of the HRI data and of the good statistics in both
observations, however,  the parameters are not well
constrained, and, moreover, they do not indicate the presence of a
component with  core radii significantly smaller than those found by
Mulchaey \& Zabludoff.  For NGC~6159, the inner core radius is of 10-20
kpc (depending on the $\beta$ parameter assumed) consistent with the
values in Mulchaey \& Zabludoff (we have rescaled their values to our
choice of \hnot).  The linear size of the smaller core radius of
IC~1262 is significantly larger than typically found in nearby systems
(40-90 kpc, depending on the assumed $\beta$).  This could be
interpreted as the ''larger" of the two Kings, since we could not probe
the innermost regions.  However, this would imply the presence of an
even larger, $third$  King model if $\beta$=1.  While we stress that we
have not tried to obtain a ``best fit model" for these data
(therefore a large range of parameter space is still unexplored), 
Fig.~\ref{iprof}
and Fig.~\ref{nprof} show that these are reasonable parameterizations
of the data.  We therefore are led to conclude that we need a
relatively large  ``central" model, larger than high resolution data of nearby
systems suggest ($\le 1-3$ kpc, Trinchieri et al. 1986, 1997b; Thomas
1986; Dahlem \& Stuhrmann 1998).    This could be due to the fact that
there systems are farther away than most galaxies probed so far
(3 kpc would be $\le 0\farcs05$ at the distance of these
galaxies).  But this also suggests that the group's potential could
itself be  composed of two components, neither of which is associated
with the central galaxy.  The presence of the arc in IC~1262 prevents us
from exploring the profile at very small radii, however even a 20 kpc
core radius would not fit the data outside $\sim 1'$, for these choices
of $\beta$.

\subsection{ The Point Source in IC~1262}

The point-like source to the NW of IC~1262 (Fig.~\ref{map-ic}) is most
likely due to an interloper, clearly visible in the optical images.
Although no spectral information is available for it, it appears to be
point-like also in red-band images, with a  magnitude m$_r \sim 16.2$
(Saglia, private communication).  We can therefore estimate an
X--ray-to-optical flux ratio, that can be compared to that of stars or
of AGNs.  If this is an AGN, we expect V-R of 0 to 0.8 (Cristiani \&
Vio 1990), therefore log(f$_x$/f$_v$) $\sim -1.3$ - $-1$ marginally
consistent with the AGN population (see nomogram in Maccacaro et al.
1988).  However, a star of K - M type could also be the optical
counterpart  of this X--ray source;  with average V-R colors of 0.8-1.1
(Allen 1973), log(f$_x$/f$_v$) $> -1$, consistent with stellar colors
(Maccacaro et al.  1988).    Spectral data are needed to properly
identify this source.

\section{The complex feature in IC~1262}
\label{thearc}

The presence of an additional feature, in the form of an arc and of a
tail, at the center of the IC~1262 source, is a peculiarity of the
X--ray emission from this galaxy.  It represents a significant 
enhancement over the immediate surrounding and it is most likely
evidence of shocked material, linked to the
galaxy and/or the group.   

The comparison between the X--ray morphology and the optical image
(Fig.~\ref{map-ic}), however, indicates a significant off-set between
the peak of the X-ray emission and IC~1262 ($\sim 20''$), 
and could further suggest an association with a $\sim 15.5$ red
magnitude point-like object (Saglia, private communication), located
just south ($\sim 5\farcs6$) of the X--ray peak.  
Although no redshift is available, it is unlikely that this
object is a group member, and therefore this association provides
no immediate explanation of this feature, since its motion
should not produce any structure in the hot gas distribution.
On the other hand, the object itself could be an  X--ray source.  In
this case, if it is an AGN or a late type star of the same kind as that
associated with the point source to the NW of the arc discussed above, we
can expect $>70$ counts (see Table~\ref{features}) from this source,
that would appear superposed on the X--ray feature.  This additional
but unrelated emission would then modify the morphology of the source
associated with the galaxy, which could either
be flatter or have a less prominent peak close to the galaxy.  
Moreover, if the arc is the result of shocked material, inhomogeneities
in the ambient medium could cause enhanced emission where 
densities are higher.   We will therefore disregard the offset position of
the peak relative to the galaxy, and assume that the whole feature is
associated with IC~1262 and/or the group, and discuss the X--ray
properties and possible origins of the shock in this context.

An estimate of the density in the arc is in principle
straightforward from X--ray data, but is in fact quite uncertain due to
a) the unknown temperature and more important b) the unknown volumes of
the structures.  While the average density in the whole source derived
in the previous section can rely on the approximation to spherical
symmetry for the volume occupied by the gas,  the third dimension in
the arc-structure is almost completely arbitrary.  For the sake of
argument, we have here assumed two possible morphologies: 1) a narrow
and long feature,  with a 30$''$ (30 kpc) cross section diameter and a 100$''$
(100 kpc) length, and 2) a thick slab 30$''$ height and 100$''$ long and deep.
We have further assumed that the feature has a plasma-like spectrum,
with an average temperature of 1 keV (the density is very weakly
dependent on  T, and therefore the uncertainty in the temperature is
not so crucial to the final values).  If this is the result of a shock,
this assumption could be wrong, since no
equilibrium situation can be expected at the site of the shock.  On the
other hand, collisional equilibrium could be a reasonable approximation
if the timescale is long enough (see later);  moreover,  with this
assumption we can obtain a rough estimate of the quantities involved,
like gas mass and density, for a very crude understanding of the
feasibility of the scenario.  We find that the average gas density in
the arc is n$_e  \sim 5.7 -\ 2.7 \times 10^{-3}$ (case 1 and 2
respectively), up to a factor of $\sim 2 \times$ higher than in the
surrounding 1$'$ (60 kpc) radius region (calculated 
from the average surface brightness derived in
Tab.~\ref{features} and applied to the area within 1$'$ radius), and
$\sim 10\times$ higher than the average density in the whole source.
The associated gas mass would be $\sim 1-2 \times 10^{10}$ M$_\odot$.

The cooling time of the gas with these average densities is $\sim 0.6-1.2
\times 10^9$ yr, which is long enough for the structure to be reasonably
stable, and also consistent with the hypothesis that it is produced by
the galaxy's motion in the medium.  The projected half length of the
structure is $<50$ kpc.  To travel such a distance in $6 \times 10^8$ years,
the galaxy needs only a velocity of $\sim 90$ km s$^{-1}$.

With equivalent reasoning, we can estimate the density and gas mass in
the tail extending to the south.  The projected size is $24 \times
84$ kpc, and we assume again the third dimension to be either of them.
The gas density ranges from n$_e  \sim 1.8-3.9\times 10^{-3}$, which
again represent a factor of $\sim 2 \times$ higher densities than in the
immediate surrounding (the region between 30 and 100 kpc radii).  

If the arc is produced by the galaxy's motion relative to the 
ambient medium, velocities $v > 500 $ to $\sim 1000$ km s$^{-1}$ 
are needed in order to produce a shock, since the emission is in
the X--ray regime, and the ambient medium is itself hot ($\sim 1-2$ keV,
by analogy with other groups' temperatures).  
We have no evidence that this galaxy is traveling at such high speed in
the group.  Wegner et al. (1999) report the spectroscopic data relative
to the brighter early-type galaxies in this group.    The group is at a
mean velocity of 10311 km/s with an estimated dispersion of $\sim$ 300
km/s, based on the early type population of 5 objects.  IC~1262 has the
lowest velocity measured in the group, with $\Delta v$ $\sim 500$ km
s$^{-1}$ from the systemic velocity.  This is at odds with its
classification as a cD galaxy, since cD galaxies are almost always
associated with the systemic velocity in groups (Zabludoff \& Mulchaey 1998),
although there are exceptions in richer clusters (Zabludoff et al. 1990).
Although caution is in order, since the velocity space is very poorly
sampled in this group, it is possible that, if the tangential component
to the velocity is also large, the motion of IC~1262 could produce a
significant ram pressure onto the ambient medium to explain the
feature.

We can actually estimate the velocity that IC~1262 needs 
to produce the density contrast derived above.  
A shock of arbitrary strength will be produced by 
$$
v_1^2 = { 2 \gamma {k T_1 \over \mu \overline{m} } \over 
(\gamma - 1 ) [{\gamma + 1\over \gamma -1}  {\rho_1 \over \rho_2} -1
]} \ \  \sim 800 \ \ {\rm km s^{-1}}
$$
where 1 and 2 indicate the upstream and downstream quantities
respectively, $\mu = 0.5$ and $\gamma = 5/3$, for a monoatomic gas,  
and the density contrast is 1/2.  
As discussed above, this is very high compared to the group's velocity
dispersion, as currently measured, and to the velocity of the
galaxy relative to the group's mean. 

The arc could also be interpreted  as the signature of a recent
merger process, that has produced a shock between the two colliding
structures.  The formation of groups and clusters is thought to go
through subsequent infalls and/or merging of external material onto an
initial, dense core (see Peebles 1980).  Shock heating
should play an important role in the dynamics and evolution of clusters
and their intergalactic medium.  Numerical simulations have shown a
variety of structures, X-ray luminosity enhancements and complex  
electron-temperature distributions as a result of the collision/merging
events (Evrard 1990; Roettiger et al. 1993;  Schindler \& M\"uller
1993; Ricker 1998; and references therein).  The evidence of a large
reservoir of hot gas in this system and its relatively regular
morphology suggest that this group has not formed too recently, since
the gas has had time to virialize.  On the other hand, merger events
could be happening at any moment in the group's evolution.  The merger
scenario would have the advantage that the velocity of each individual
colliding structure does not need to be so extreme; however, there is
at the present time no evidence of merging activity in this system.
Detailed optical studies of the field, coupled with a better
understanding of the hot gas dynamics, would be needed to obtain
independent evidence of a recent merger.

\subsection{Shocks in groups. }

X--ray observations of groups have provided very limited evidence of
shocks, although the X--ray morphologies of a few galaxies in them
clearly point to some interaction with the surrounding medium.  Most of
the evidence is in the form of clear tails and/or distorted envelopes
(for example M86, Forman et al.  1979, Rangarajan et al. 1995;
NGC~7619, Trinchieri et al. 1997a; NGC~4472, Irwin \& Sarazin 1996;
NGC~5044, David et al. 1994), that have been interpreted as wakes from
the ram pressure (partial) stripping as the galaxies move in the dense
medium.  The emission associated with IC~1262 would instead trace the
bow-shock, but unlike NGC~4472 (Irwin \& Sarazin 1996), without any
visible tail behind the shock.  

A feature with a possibly  more similar morphology to the arc in
IC~1262 was found in Stephan's Quintet, where a narrow (although unbent)
X-ray source is observed between two member galaxies (NGC~7319 and
NGC~7318b), coincident with radio continuum and H$\alpha$/[NII]
emission (van der Hulst \& Rots 1981, Moles et al. 1997), and
interpreted as the result of  shock-heated material in the complex
dynamics of the system (Pietsch et al. 1997, Sulentic 1999).

However, the characteristics of both the two systems and the two shocks
are rather different.  Stephan's Quintet is a spiral-rich compact
group, displays strong signs of interaction at many wavelengths, and
there is no compelling evidence of a significant amount of hot,
intergalactic medium, although current limits need to be refined with
more sensitive X--ray observations (see Pietsch et al.  1997).  The
X--ray source is a clearly defined, intragalaxy structure that
dominates the emission from the group, although the luminosity
associated with it is not as high as in IC~1262 (L$_x \sim 2 \times
10^{41}$ erg s$^{-1}$).  IC~1262 is in contrast composed of mostly
early type galaxies, has a smaller velocity dispersion, no evidence (up
to now!) of recent intruders, but contains a large amount of hot
intergalactic medium.  The arc has a high X--ray luminosity, but does
not dominate the emission from the system, and in fact represents a
relatively minor perturbation in the diffuse hot gas emission of the
group.  However, these examples point to the necessity of a detailed
analysis of the morphology of the X-ray emission in groups of galaxies,
even in those with a clear, extended hot gas component, in order to
fully understand the characteristics of the gas in the gravitational
potential and to properly assess its contribution and the presence of
perturbations that might otherwise go undetected.   A more
comprehensive assessment of the information provided by the
intergalactic medium is crucial for a proper understanding of the
formation and evolution of groups of galaxies.

\section{Conclusions}

We have presented high spatial resolution X--ray data of two galaxies unusual
in their higher than expected X--ray emission relative to the optical
brightness.   It is likely that the emission detected is due to small
groups of galaxies that appear to be associated with the primary
optical identification.  Redshift information confirms in fact the
existence of a small group of early type galaxies around IC~1262.  The
reality of a group around NGC~6159  is also likely, but
needs spectroscopic confirmation.

The emission appears irregular and not azimuthally symmetric at large
radii.  However, fits to the azimuthally averaged surface brightness
profiles are possible, and suggest the presence of two X--ray
components.   In spite of the high spatial resolution of the
X--ray data, we do not find evidence of a small central component
similar to those observed in high resolution images of more nearby
objects.  While failure to detect this could be due to the large
distances of these systems, the radial profiles of these objects could 
suggest a complex shape of the group's potential. 

We found a peculiar feature in the X--ray emission of IC~1262, in the
form of an arc, probably the result of shocked material either caused
by peculiar motions within the group's potential, or resulting from a
recent merger event.  To our knowledge, this is the first such feature
observed in a poor group of mostly early-type galaxies.  A confirmation
of the proposed interpretation for the peculiar structure would come
from the study of the temperature distribution of the group's gas,
which should be hotter in the arc than in the surrounding region.
Unfortunately, no temperature information can be derived from the HRI
data.

\appendix
\section{Other X--ray sources in the IC~1262 and NGC~6159 fields} 
\label{appen}

\begin{table*}
\caption{X-ray properties of the sources in the  IC~1262 and NGC~6159 field and 
their proposed identifications 
}
\label{coin}
{
\begin{flushleft}
\begin{tabular}{lrcclrrc}
\hline
\noalign{\smallskip}
ROSAT name &  X--ray position &Net&
X--ray&
\multicolumn{1}{c}{Identification} &$\Delta$X-O
&Notes \\
\noalign{\smallskip}
(RX J)  & \multicolumn{1}{c}{(2000.0)}
 &counts& \multicolumn{1}{c}{flux}&
& \multicolumn{1}{c}{(\arcsec )}&           \\
\noalign{\smallskip}
\multicolumn{1}{c}{(1)} &
\multicolumn{1}{c}{(2) }&
\multicolumn{1}{c}{(3)} &
\multicolumn{1}{c}{(4)} &
\multicolumn{1}{c}{(5)} &
\multicolumn{1}{c}{(6)} &
\multicolumn{1}{c}{(7)} \\
\noalign{\smallskip}
\hline
\noalign{\smallskip}

\multicolumn{7}{c}{IC~1262 field} \\
\hline
\noalign{\smallskip}
173222.2+433949 &17:32:22.2 43:39:49.1& 249$\pm$18 &3.9$\times 10^{-13}$&
U1275\_09527890 
& $0\farcs6$&1 \\ 
173258.7+434613& 17:32:58.7 43:46:13.5& 136$\pm$14 &2.1$\times 10^{-13}$&
U1275\_09534024 
& $2\farcs4$&1 \\ 
		
173334.0+434757 &17:33:34.0 43:47:57.2& 154$\pm$15 &2.4$\times 10^{-13}$&
U1275\_09539872 
 &$1\farcs9$&1 \\ 
&&&&U1275\_09539788 &$1\farcs9$&1 \\ 
		
\noalign{\smallskip}
\hline
\multicolumn{7}{c}{NGC~6159 field} \\
\hline
\noalign{\smallskip}
162658.2+424449 & 16:26:59.2 42:44:49.3& 49$\pm$12  & 3.0$\times 10^{-14}$
&U1275\_09047679 
&$2\farcs2$&1 \\ 
162711.3+424111 & 16:27:11.3 42:41:11.5& 54$\pm$12 & 3.3$\times 10^{-14}$
&U1275\_09048889 
&$2\farcs6$&1 \\ 
&&&&U1275\_09048880 
&$2\farcs9$&1 \\ 
&&& &GSC 0306901509 
&$2\farcs9$&2 \\ 
162713.8+424142& 16:27:13.8 42:41:42.7&  61$\pm$13  & 3.8$\times 10^{-14}$
&U1275\_09049154 
&$2\farcs5$&1 \\ 
(162736.4+424455& 16:27:36.4 42:44:55.2& 54$\pm$13  & 3.3$\times 10^{-14}$
&U1275\_09051638 
& $11\farcs7$& 1) \\ 
\noalign{\smallskip}
\hline
\noalign{\smallskip}
\hline

\end{tabular}
\end{flushleft}
Notes: \\ 
unabsorbed fluxes are in the 0.1--2.0 keV band, and are derived from the HRI count rates,
assuming a 5~keV thermal bremsstrahlung spectrum and line-of-sight 
for Galactic absorption\\
(1) Identified with a source in the ESO/ST-ECF USNO-A1.0 \\ 
(2) Identified with a source in the Guide Star catalog \\
}
\end{table*}

Four point-like sources are detected around NGC~6159 and two sources
around IC~1262 in these HRI observations.   An additional one is
clearly visible close to IC~1262 (see Fig.~\ref{map-ic}).
Table~\ref{coin} shows their characteristics.  For each X--ray source
in the field, we list the closest optical candidate(s) from either the
USNO catalog or the Guide Star Catalog (obtained from searches in the
Online USNO-A1.0 Catalogue Server and HST Guide Star Catalogue (GSC)
Server at the ESO/ST-ECF Archive).  The next optical candidate is $>
10''$ away from the X--ray source.  As clearly visible, the mean offset
of the X--ray and optical positions (listed in col. 6 of
Table~\ref{coin}) is very small, which ensures that our absolute
coordinate system for the X--ray images are accurate.  The closest
optical counterpart to source RXJ162736.4+424455 is at 12$''$.  Since
this $\Delta $X-O is significantly larger than all others, the optical
source might not be the correct identification for the X--ray source.
Most of the sources are very faint and have no other identification in
published catalogs.

\begin{acknowledgements} 
This work has received partial financial support from the Italian Space
Agency.  We thank the referee, Dr. J. Bregman,  Dr. Piero Rosati for a careful
reading of the manuscript and Dr. Dieter Breitschwerdt for his patience in
discussing shocks.  GT thanks Prof. Tr\"umper and the MPE for hospitality while
part of this work was done.   Dr. Roberto Saglia kindly provided us with
useful information prior and beyond publication.  This research has
made use of the NASA/IPAC Extragalactic Database (NED) which is operated by the
Jet Propulsion Laboratory, California Institute of Technology, under contract
with the National Aeronautics and Space Administration.

\end{acknowledgements}

\end{document}